\documentclass[aps,pre,onecolumn,superscriptaddress]{revtex4}  
\usepackage{graphicx,epsfig}  
\usepackage{dcolumn}   
\usepackage{bm}        
\usepackage{amssymb}   
\usepackage{color}
\usepackage{mathtools}
\usepackage{esvect}
\usepackage{multirow}
\usepackage[outercaption]{sidecap}
\usepackage{natbib}

\begin{document}

\title{Electromagnetic   wave   transparency  of X mode  in strongly magnetized plasma }

\author{Devshree Mandal}
\affiliation{Institute for Plasma Research, HBNI, Bhat, Gandhinagar 382428, India}
\affiliation{Homi Bhabha National Institute, Mumbai, 400094, India}
\author{Ayushi Vashistha}
\affiliation{Institute for Plasma Research, HBNI, Bhat, Gandhinagar 382428, India}
\affiliation{Homi Bhabha National Institute, Mumbai, 400094, India}

\author{Amita Das}
\affiliation{Department of Physics, Indian Institute of Technology Delhi, Hauz Khas, New Delhi 110016, India}

\begin{abstract}
 An Electromagnetic (EM) pulse falling on a plasma medium from vacuum can either reflect or propagate inside the plasma depending on whether 
it is overdense or underdense. In a magnetised plasma, however, there are usually several pass and stop bands for the EM wave depending 
on the orientation  of the magnetic field with respect to the propagation direction.  The EM wave while propagating in a plasma 
can excite  electrostatic disturbances  in the plasma \cite{Vashistha_2020, Vashistha_2021}.  In this work Particle - In - Cell simulations have  been carried out to illustrate the 
complete transparency of the EM wave propagation inside a strongly magnetised plasma. 
The external magnetic field  is chosen to be perpendicular to both the wave propagation direction and the electric field of the EM wave, which is the  X mode configuration.  
 Despite the presence of charged electron and ion species the plasma medium behaves like a vacuum. 
  The observation is understood with the help of particle drifts. It is shown that though the two particle species move under the influence of EM fields 
  their motion does not lead to  any charge or current source to 
  alter the dispersion relation of the EM wave propagating in the medium. Furthermore, it is also shown that 
  the stop band for EM wave in this regime 
  shrinks to a zero width as both the resonance and cut-off points approach each other.  Thus transparency to the EM radiation in such a strongly 
  magnetised case appears to be a norm.  This may have important implications in astrophysical scenarios.  
  For instance, the plasma surrounding objects like 
  pulsars and magnetars  is often threaded with strong magnetic fields.

\end{abstract}

\maketitle

 \section{Introduction}

In astrophysical plasmas, a wide range of magnetic fields exist e.g. in galaxy clusters, shock
formation in gamma ray bursts, magnetospheres of neutron stars \cite{Medvedev_2006, Vikhlinin_2001, 2003A&A...412..373V,mag1}. Several instabilities also lead to the generation of intense magnetic field in the astrophysical plasmas e.g. magnetic field generation via kelvin-Helmholtz instability, counter-streaming electron flows etc \cite{Medvedev_2006a,Alves_2012,P_Alves_2014}. Thus   plasmas associated with astrophysical objects are often  threaded with   magnetic fields. These magnetic fields 
can be very strong  in some cases.  For instance,  near pulsars and magnetars they could be of the order of Giga Teslas \cite{magnetar, Parent_2011, Camilo_2000}. 
It is thus possible that electromagnetic wave would encounter a strongly magnetized plasma in such astrophysical scenarios.

 Such scenarios involve understanding the interaction of EM waves with matter which has been covered with tremendous interest for many decades now \cite{Kaw_1969, Dawson_oberman, kaw_rmp}. The principle mechanisms in plasmas depend on  EM wave frequency and plasma permittivity. Plasma permittivity can be altered by suitably choosing the plasma density and applied magnetic field. The free charges and their associated currents in the plasma medium act as sources and influence the plasma dielectric constant making it different from  vacuum. Propagation of EM wave through strongly magnetized plasma sources, therefore,  needs to be understood. 

The issue of EM wave achieving complete transparency  is important and has been considered earlier in many contexts. 
An attempt to achieve transparency using strong fields generated by intense $fs$ laser pulse has been studied by \cite{gizziPRE, gizziPRL, RT1, RT2, RT3}. Total transmission was observed when  $30 fs$ laser pulse of intensity $3\times10^{18} Wcm^{-2}$ passes through $0.1 \mu m$ plastic foil targets (\cite{gizziPRL}). This mechanism is operative when the target width is much smaller than the laser wavelength \cite{gizziPRE}. However, for thick targets, relativistic laser would lead to excitation of coherent structures and/or instabilities leading to turbulence in the system\cite{mandal_vashistha_das_2020,chatterjee2017}. 

External magnetic fields have been applied in several contexts  to achieve transparency. The dressing of resonance states for RHCP( Right Hand Circularly Polarized) waves  with the combination of axial and wiggler magnetic field is attempted to acquire a window of transparency in the opaque magnetized plasma for the EM wave \cite{PRL2012}. Another technique to seek transparency is by employing  pump electromagnetic wave to transmit the probe wave (\cite{EIT1,EIT2,EIT3,EIT4}). This method is analogous to a quantum mechanical phenomena known as EIT(Electromagnetically Induced Transparency). In this phenomena, an electromagnetic wave is made to propagate in an normally opaque medium in presence of powerful secondary EM wave. This is possible due to the destructive interference between several energy levels connecting the ground and excited states of the atom. This method is heavily used in non-linear optics to manipulate the energy levels of atomic states or slow down the waves\cite{Q1,Q2,Q3}. In plasma, the use of pump and probe EM wave is used to make plasma transparent to RHCP wave. But this study is limited to propagation of RHCP waves along the magnetic field lines and generating an additional wiggler magnetic field  is a complication from application viewpoint. More importantly these methods focus on R-wave mode where propagation is along the magnetic field, while L-wave and X-mode propagation geometries are yet to be explored.\\

We have carried out Particle - In - Cell simulations to study the propagation of EM wave in a  strongly magnetized plasma for which both the electron and ion species are 
strongly magnetized. An interesting observation of complete transparency of the plasma medium is observed for the propagation of EM wave in the X mode configuration. 
The width of the stop band of the X mode reduces to   zero  and a completely transparent propagation of EM wave is observed.  
  For X-mode, dispersion curve is given in fig.\ref{dispersion}, there are two stop bands ($\omega_{LH}-\omega_{L}$ $\&$ $\omega_{UH}-\omega_{R}$, where $LH,UH,L,R$ stands for Lower Hybrid, Upper Hybrid, Left Hand Cutoff, Right Hand Cutoff respectively). Whenever laser/EM wave frequency lying on these bands is incident on plasma, it generates a shielding electric field in response to avoid penetration of EM wave inside it. The propagation of EM wave in plasma thus depends on the intensity and/or frequency of the incident EM wave. We show that when the strong external magnetic  field dominates the motion of both charged species, \emph{i.e.} $ \omega_{ce} > \omega_{ci} > \omega_{l}$, or  it strongly magnetizes electrons but the perturbations are at faster time scales than that of ions    $ \omega_{ce} > \omega_{l} > \omega_{LHC}$, the electromagnetic wave propagates undisturbed inside plasma. The inequalities gives an insight to plasma in terms of anisotropy and modification in its collective behaviour that gets affected by  introduction of magnetic field.  Such caveats are vital for absorption phenomena as well.

  A strong  magnetic field in which the two charged species remain closely tied  to the  magnetic field at the EM wave frequency, 
  ensures that they do not provide the plasma a chance to respond to the incoming laser/ EM wave. We show that as the strength of magnetic field increases, the magnitude of self generated fields in bulk plasma due to interaction of laser with plasma decreases and the  EM wave propagation speed inside the plasma approaches the speed of light $c$. The plasma medium thus acts 
  as a transparent to the incoming EM wave. 
 These observations  would have important implications in astrophysics. For instance, the  strength of magnetic field in neutron stars and magnetars goes upto Giga-Tesla where this study is expected to be  applicable. \\ 
 
 The paper has been  arranged as follows:  Section II describes our PIC simulation, Section III contains the analysis and the discussion   under different sub-sections and Sec.IV provides concluding remarks.

	\section{ Simulation Details}
	
	\begin{table}
	\centering
		\caption{Details of various simulation runs used in this study based on different regions of X-mode shown in fig. \ref{dispersion}}
		\begin{tabular}{|p{4cm}||p{3.5cm}||p{3cm}||p{1cm}||p{1cm}||p{1cm}|}
			\hline
			{}&\textcolor{red}{ $B_0$ criteria I ($\omega_{ce}> \omega_{l}>\omega_{ci}$)}& \textcolor{red}{$B_0$ criteria II ($\omega_{ce}>\omega_{ci}>\omega_{l}$)}& 	\textcolor{red}{$\omega_l$ (in $\omega_{pe}$) }& \textcolor{red}{$a_0$}& \textcolor{red}{$E_0$}\\
			
			\hline
			\hline
			
			Region I (0-$\omega_{LH}$)& 3, 8& 20&0.05&0.6&0.03\\
			\hline
			Region II ($\omega_{LH}-\omega_{L}$)& 3, 8 &20, 40&0.2&0.15&0.03\\
			\hline
			Region III ($\omega_{L}-\omega_{UH}$)& (0.25,0.5 \textcolor{red}{ $\omega_{ce}<\omega_l$}),3, 8&40&0.5&0.06&0.03 \\
			\hline
			
			\hline	
			
		\end{tabular}
	\end{table}

	 We have carried out series of one dimensional (along $\hat{x}$) PIC simulations in X-mode configurations using OSIRIS-4.0 \cite{hemker,Fonseca2002,osiris}. For X-mode configuration, uniform external magnetic field($B_0(m_e\omega_{pe}c/e)$) has been applied in $\hat{z}$ direction.  A uniform plasma density comprising of electrons and ions has been considered. Ion mass is taken to be 100 times mass of electrons ($m_i=100m_e$) for faster computation. Plasma boundary extends from $x=850 c/\omega_{pe}$ to $x=2000c/\omega_{pe}$ whereas total length of simulation box is $3000 c/\omega_{pe}$. Boundary condition for particles as well as fields are absorbing. A  p-polarized, plane $ CO_2 $ laser pulse is incident normally at plasma from the left boundary. Laser is  propagating along $ \hat{x} $ with its spatial profile centered at x= $450 c/\omega_{pe}$ and ranging from x= $0 $  to $800 c/\omega_{pe}$. We also want to clarify that this work focuses on proof of concept so the mechanism presented in this paper depends on the magnetisation of the charge species with respect to the incoming EM pulse frequency. We have carried out a parametric study on magnetic fields such that broadly they follow either criteria I ($ \omega_{ce} > \omega_{l} > \omega_{ci}$) or criteria II ($ \omega_{ce} > \omega_{ci} > \omega_{l}$). This parametric study has been done with laser pulse of intensity lying in non-relativistic regime such that amplitude of laser electric field($E_0=0.03(m_e \omega_{pe}c/e)$) is constant for all runs. This has been done to avoid other relativistic mechanism to play a role. A schematic of simulation geometry has been shown in Fig.\ref{schematic}.  Different laser frequency maintaining criteria I and criteria II has also been chosen according to regional frequency which is explained more elaborately in next section.  A tabular form of simulation parameters is given in table I.

	\section{Results}	
	\subsection{Theory and Analytical assessment }
It is well known that when EM wave is propagating perpendicular to external magnetic field, plasma supports two kinds of waves, O-mode (ordinary wave) and extraordinary mode (X-mode). O-mode is independent of applied magnetic field  (ordinary wave).
The general dispersion relation for   perpendicular propagation in cold plasma ($\vec{k}\perp\vec{B}$) is given by the matrix,
\begin{equation}
\begin{bmatrix}
S & -iD & 0\\
iD & S-n^2 & 0\\
0 &0 & P-n^2
\end{bmatrix} 
\begin{bmatrix}
E_x\\
E_y\\
E_z
\end{bmatrix}=0
\end{equation}
where, $ S=\frac{1}{2}\left(R+L\right)$, $D=\frac{1}{2}\left(R-L\right)$, $P=1-\frac{\omega^2_{p}}{\omega^2}$

\begin{equation}
	\label{Rmode}
	 R=1- \frac{\omega_{pe}^2 + \omega_{pi}^2}{(\omega + \omega_{ci})(\omega- \omega_{ce})}
	\end{equation}
	
\begin{equation}
\label{Lmode}
L=1- \frac{\omega_{pe}^2 + \omega_{pi}^2}{(\omega - \omega_{ci})(\omega+ \omega_{ce})}
\end{equation}

	The X-mode has R and L  mode having cut-offs at $\omega_R$ and $\omega_L$ respectively. $\omega_R$ and $\omega_L$ are given as follows:

\begin{equation}
\label{wr}
\omega_{R,L}= [\omega_{pe}^2+\omega_{pi}^2+(\omega_{ci}+\omega_{ce})^2/4]^{1/2} \mp(\omega_{ci}-\omega_{ce})/2
\end{equation}	
Dispersion curve of X-mode is shown in fig.\ref{dispersion}. We have classified dipersion curve into three regions depending on the dominant role played by the species. Region I is dominated by dynamics of ions and Region III for electrons. Region II is stop band as it lies between $\omega_{LH}$ (resonance point) and $\omega_{L}$ (cut-off point).

Dispersion relation for X-mode is obtained
\begin{equation}
n^2= \frac{RL}{S}
\end{equation}
Principal resonance occur when $S \rightarrow 0$
\begin{equation}
\label{analytic}
\omega^4-\left(\omega_{pe}^2+\omega_{pi}^2+\omega_{ce}^2+\omega_{ci}^2\right)\omega^2+\omega_{ci}^2\omega_{ce}^2+\omega_{pe}^2\omega_{ci}^2+\omega_{pi}^2\omega_{ce}^2=0
\end{equation}
 This is a bi-quadratic equation, it's lower end solution is  plotted as function of applied magnetic field in fig.\ref{analytic1}. As can be seen from the figure, when $B_0<10$(at $B_0=10, \omega_{ci}=\omega_{pi}$) it falls in criteria I and solution of eq. \ref{analytic} matches perfectly with reduced expression of $\omega_{LH}$. At higher magnetic fields, $\omega_{LH}$ saturates at $\omega_{pi}$ while solution of eq. \ref{analytic} approaches left hand cut off ( $\omega_{L}$) asymptotically which concludes that at this higher magnetic field the resonance point and cut off approach each other thus effectively reduce the width of the stop band.  This was checked by simulation as well for frequency parameter lying in region II(i.e. stop band). Under criteria I, laser reflected back. On the other hand, under criteria II laser pulse was able to propagate through the plasma. This was possible due to effective reduction of stop band and resonance point lying well above EM frequency. So, effectively this case does not lie in region II but in region I.
 
 In Fig.\ref{xmode_compare}, we have plotted the dispersion curve for X-mode in two criteria. As one can observe from the left subplot that in criteria I, all the regions are well separated while in the criteria II stop band has shrunk. Moreover, the dispersion follows $\omega=k$(as $c=1$). Therefore, it rules out any other mode excitation when $\omega_{ce} > \omega_{ci} > \omega_{l}$. \\
 
  To summarize our propagation characteristics according to their region of dispersion curve in fig. \ref{dispersion}. is given in table II.
  \begin{table}
  \centering
		\caption{Details of propagation characteristics observed in various simualtion runs for different regions of dispersion curve}
		\begin{tabular}{|p{2cm}||p{2cm}||p{2cm}|}
			\hline
			{}&\textcolor{red}{ criteria I }& \textcolor{red}{ criteria II}\\
			\hline
			\hline
			Region I & LH& Transparent\\
			\hline
			Region II & Stop band&Transparent\\
			\hline
			Region III & Transparent& N.A\\
			\hline		
			\hline	
			
		\end{tabular}
	\end{table}	
	Detail quantitative analysis to calculate absorption, reflection, transmission coefficients has also been done which is presented here in tabulated form (Table III).
	 \begin{table}
	 
		\caption{A comparison of various coefficients with different external magnetic field}
		\begin{center}

		\begin{tabular}{|p{2cm}||p{1cm}||p{1.5cm}||p{1.5cm}||p{1.5cm}|}
			\hline
			{}&\textcolor{red}{ $B_0$}& \textcolor{red}{R}& \textcolor{red}{T}& \textcolor{red}{A}\\
			\hline
			\hline
			\multirow{3}{*} {Region I}  &3&0.67&8.9$\times10^{-9}$&0.32\\
			
			&8&0.0663&0.872&0.0246\\
			
			&20&0.0035&0.9931&0.02\\
			\hline
			\multirow{4}{*} {Region II}  &3&0.99&7.95$\times10^{-9}$&1.9$\times10^{-5}$\\
			
			&8&0.998&$8\times10^{-7}$&0.0036\\
			
			&20&0.0089&0.9823&0.0047\\
						&40&5.6$\times10^{-4}$&0.9989&0.0018\\

			\hline
			\multirow{3}{*} {Region III}  &3&0.013&0.9742&0.003\\
			
			&8&0.006&0.9988&0.0019\\
			
			&40&0.0035&0.9931&0.02\\
			\hline
			\hline

		\end{tabular}
		\end{center}
		\label{coefficient_table}
	\end{table}
	One comment should be made here about another solution in the upper end of frequency scale, it was found that at high magnetic field region $\omega_{ce}$ dominates all modes and cut off points and hence they merge very well. Plot of exact solution of eq.\ref{analytic} as function of $B_0$ applied is given in fig.\ref{analytic2}. As one can observe here gap between $\omega_R$ and $\omega_{UH}$ is very thin and at high magnetic fields they also merge indicating that stop band at upper frequency also vanishes with application of strong  magnetic field.  
	\subsection{$\omega$ and $\vec{k}$ Analysis }

 In any dispersive medium, as the  refractive index of media change spatially,  the  frequency of the EM  wave remains same while its wavelength 
 suffers a change. In this section we calculate  the modified $k$ and the phase velocity of incident laser pulse. We observe that by varying ambient magnetic field, phase velocity of laser pulse also changes (approaches velocity of light in vacuum,$c$) while decreasing the perturbations in the plasma.

 Fig.\ref{ExEy} shows a comparison of all three cases lying in Region I. Initially (at t=0), electric field due to laser is present in the system. In case(A) ($B_0=0$), the laser interacts with plasma and gets reflected back from the plasma surface. However, for case(B) ($B_0 = 3$, satisfying the condition $ \omega_{ce} > \omega_{l} > \omega_{ci}$), there are certain modes generated in plasma and as a result we observe a finite magnitude of $E_x$ in the system. $E_x$ that get generated in plasma have higher magnitude than $E_y$.  On the other hand, in case(C) ($B_0=20$, satisfying the condition $ \omega_{ce} > \omega_{ci} > \omega_{l}$), the plasma seems to be completely undisturbed by the laser as pulse freely propagates inside it without creating any perturbations in the medium and goes into vacuum space in the right side. The transparency induced in plasma on applying external magnetic field is the key observation of this work. Plasma density plots show that in case(A), plasma density at the interface is modified, on the contrary, for case(B), density perturbations are present in the bulk plasma as well. Ion density fluctuates more than electrons which propagates in longitudinal direction as can be seen at later times in fig.\ref{neni}. For case(C), there being  density perturbations that can be seen at $t =500$ is due to laser field \emph{i.e.} electrons and ions fluctuates with same amplitude, justifying our observation that plasma remains undisturbed via interaction with laser in this case[fig. \ref{neni}].  \\

  \begin{table}
  	\caption{Velocity of EM wave (normalised to c) in plasma with changing magnetic field}
  	\begin{center}

  		\begin{tabular}{|p{0.75cm}||p{2.5cm}||p{2.5cm}||p{2.5cm}|}
  			
  			\hline
  			\textcolor{red}{$B_0$ }& \textcolor{red}{Velocity from method I}& 	\textcolor{red}{Modified $K_x$}& \textcolor{red}{Velocity from method II}\\
  			\hline	
  			8&0.78&0.2518&0.79\\
  			\hline
  			15& 0.87 &0.228&0.88\\
  			\hline
  			20&0.92&0.2137&0.94\\
  			\hline
  			40&0.96&0.2030&0.98\\
  			\hline

  		\end{tabular}
  	\end{center}
  		\label{v_table}
  \end{table}
 
 Fig. \ref{fft_t} shows the FFT of $B_z$ of laser with respect to time for four different value of magnetic field, where transparency has been induced.  It can be seen that the frequency of laser($\omega_l=0.2$) does not change while  propagating inside plasma (we show FFT of $B_z$ with time at two different values of x in the bulk plasma and obtain the same peak.). However, $k_x$ of EM wave gets modified on propagation inside plasma (figure \ref{fft_x}). The shift from the initial value of $k_x$ decreases on increasing applied magnetic field. We calculate the modified velocity of EM wave in plasma by  peak frequency of the wave from the FFT and  modified $k_x$ value (method II in table \ref{v_table}) and found that velocity of the wave inside plasma approaches to $ c$ on increasing applied magnetic field ( Table \ref{v_table}). In table \ref{v_table}, we calculate velocity by two methods. In Method I, we choose a point on the waveform and calculate the time taken by that point to cover a particular distance and method II includes calculation of velocity by modification in $k_x$. So we conclude that strong magnetization can stop pulse modification while pulse waveform is propagating through plasma media. 
 
 \subsection{Reversible and Irreversible exchange of energy}
 In this study, we observed that depending on region and criteria, the laser energy exchange is either reversible or irreversible. 
 As one can notice from table\ref{coefficient_table}, in region I criteria I there is significant absorption and this region is well explored in ref\cite{Vashistha_2020}. From these studies we know that in region I criteria I energy is dominantly coupled to ions and  this coupling process is irreversible. While in region I criteria II, due to tansparency, energy transfer is observe to be reversible. As when laser is present in the plasma, electrons and ions oscillate due to oscillating electric field and when the field passes through, they come to rest.\\
 In region II criteria I, laser reflects back due to formation of shielding fields so there's no exchange of energy altogether. On the other hand region II criteria II is effectively Region I criteria II so there's similar exchange of energy which is reversible.\\ 
 In region III criteria I, we observe $97 \%$ transparency and reversible exchange of energy. This is quite different from other two regions, the reason behind this is simple. As the time scales of region III are same as electrons, with $B_0=3$ electrons are strongly magnetized. So that's why laser is not able to couple its energy into electron effectively. To couple laser energy into electron irreversibly one has to weakly magnetise the electron and that can be achieved by ensuring another inequality \emph{i.e.} $\omega_{ce}<\omega_{l}<\omega_{L}$. When we simulated with this condition by taking $B_0=0.25$, we observe $10.2\%$ absorption into electrons and about $2.2\%$ energy to ions irreversibly while $88 \% $ of laser pulse was reflected back. \\
 We concluded that when the species are tightly bound to external magnetic field, they are not able to take energy from EM pulse irreversibly. That's why in region I and criteria II when both the species were tightly magnetized to external magnetic field they were unable to couple their motion to laser pulse and that's how pulse was transparent in this medium. In region III where it is in propagating region when electrons were tightly bounded we observe transparency for similar reason. One can argue that ions are not magnetized in this condition but this region's time scales are fast so only electron motion is important here. 
 
  \subsection{Drift Velocity Comparison in Different Regions}
  In this section, we will provide proof on how charge species are magnetized or unmagnetized in different regimes and their analytic estimates are compared to their simulation values. 
		Under the effect of oscillating electric field and external magnetic field, the longitudinal drift can be written by eq.\ref{EXB}
		\begin{equation} 
\label{EXB}
\vec{V}_{\vec{E} \times  \vec{B}}(t) = \frac{ \omega_{cs}^2}{ \omega_{cs}^2 - \omega_{l}^2} \frac{ \vec{E}(t) \times \vec{B}}{B^2}
\end{equation}
  	
  	Here, the suffix $s = e, i $ represents the electron and ion species respectively.
		Like in Criteria II, both ions and electrons are strongly magnetised and the dynamics are governed by the Lorentz force i.e. eq.\ref{EXB}. And as longitudinal velocity  is independent of mass of the species for this criteria, there's no possibility of charge separation such that there are no shielding fields to restrict the EM pulse propagation. A comparative analysis has been done between analytical drift given in eq.\ref{EXB} and numerical longitudinal drift experienced by both ion and electron in subplot(B) of fig. \ref{comparison} and it can be observed that there's no velocity difference between species which results in no net charge density separation and hence EM wave is able to propagate unhindered. 
		In Criteria I, electrons follow eq.\ref{EXB} while ion motion is governed by electric fields as they are unmagnetized. This is shown in subplot (A) and (C) of fig.\ref{comparison}. We observe that they match well. 
		When $\omega_{ce}<\omega_{l}$(named as Criteria 0), both species are unmagnetized and they follow the longitudinal electric field (see subplot (D) of fig.\ref{comparison}). 
		
\section{conclusion}
A detail PIC simulation has been carried out by us to show 
 complete transparency of EM wave radiation through a plasma in the presence of strong 
ambient field. The strength of the magnetic field has to be  strong enough to  elicit magnetised response from  both  electron and ion species  at the EM wave frequency. 

The effect is understood by realising that at such strong magnetic field the 
  the hybrid modes essentially vanish and in that regime resonance point and cut-off points approach each other resulting in  reduction  and eventually  disappearance of the width of the stop band. Thus, the EM wave freely propagates inside the plasma medium for any choice of frequency in this particular regime. It has also been 
  shown that the particle drifts are identical for both electron and ion species at strong magnetic field. Thus,  
  plasma provides for no charge and current sources in the EM wave propagation. The electrostatic perturbations, 
( which lead  to irreversible transfer of EM wave energy to plasma)    normally occur due to differential drift 
 of the two species. In this case the drifts being identical there is no conversion to electrostatic perturbations of the EM wave. 
 
 We feel that these observations  will have important significance in the context of astrophysical plasma near 
 pulsar and magnetars where the magnetic field is quite strong and would elicit  magnetised  plasma response 
 for typical EM frequencies of interest. 

\section*{Acknowledgements}

The authors would like to acknowledge the OSIRIS Consortium, consisting of UCLA ans IST(Lisbon, Portugal) for providing access to the OSIRIS4.0 framework which is the  work supported by NSF ACI-1339893. AD would like to acknowledge her  J. C. Bose fellowship grant JCB/2017/000055 and the CRG/2018/000624 grant of DST for the work. The simulations for the work described in this paper were performed on Uday, an IPR Linux cluster.\\


\bibliography{mag_trans_ref}





\begin{widetext}

\begin{figure*}
		\includegraphics[width=\textwidth]{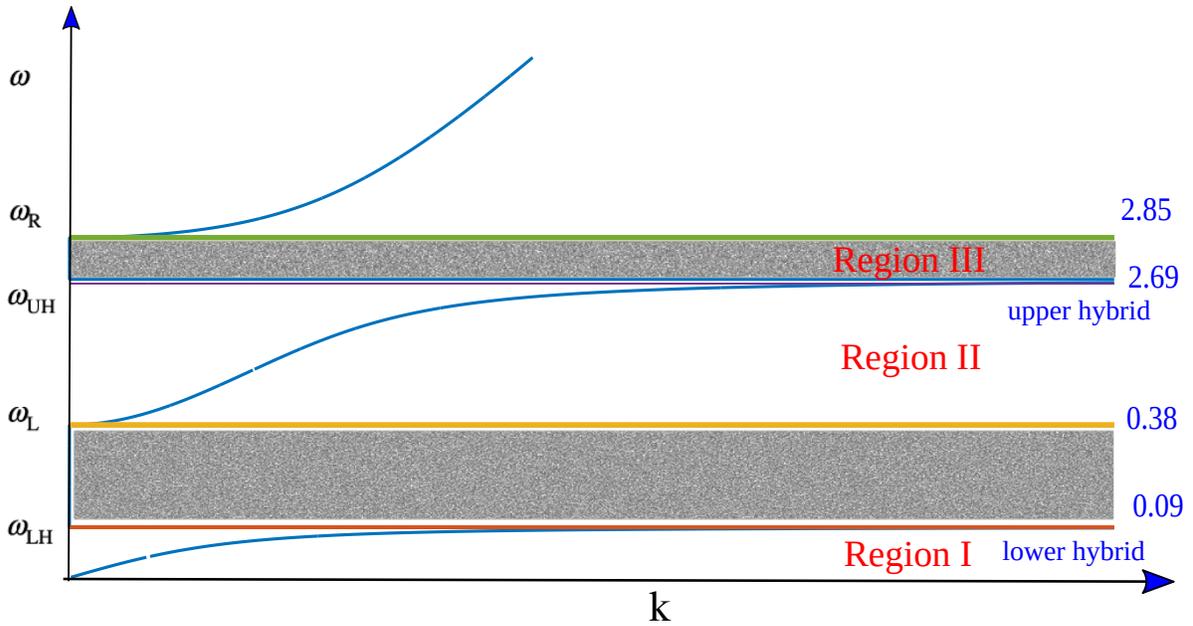}

		\caption{Dispersion of X-mode when $\vec{k} \perp \vec{B_0}$, classified into three regions. The numbers on the right hand side correspond to resonance and cutoff frequencies for $m_i=100m_e$ in normalized units. }
		
		\label{dispersion}
		
	\end{figure*}
	
\begin{figure*}
	\includegraphics[width=\textwidth]{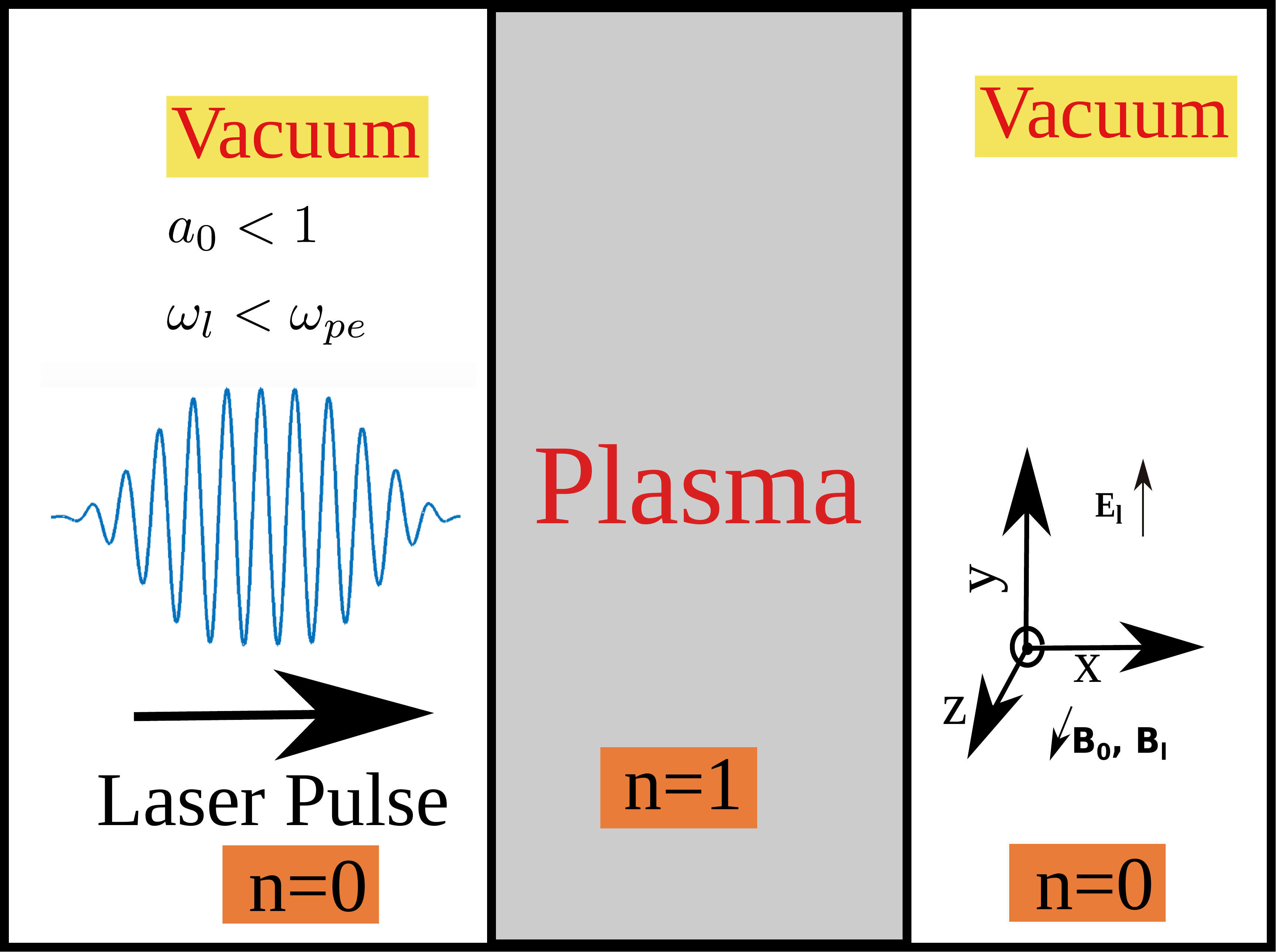}

	\caption{Schematic of simulation geometry. Plasma slab is placed in vacuum, direction of external Magnetic field($B_0$) is along the same direction as that of the  magnetic field of laser($B_l$)   \emph{i.e.} $\hat{z}$. Electric field of laser is along $\hat{y}$ direction. }
	
	\label{schematic}
	
\end{figure*}

	\begin{figure*}
		\includegraphics[width=\textwidth]{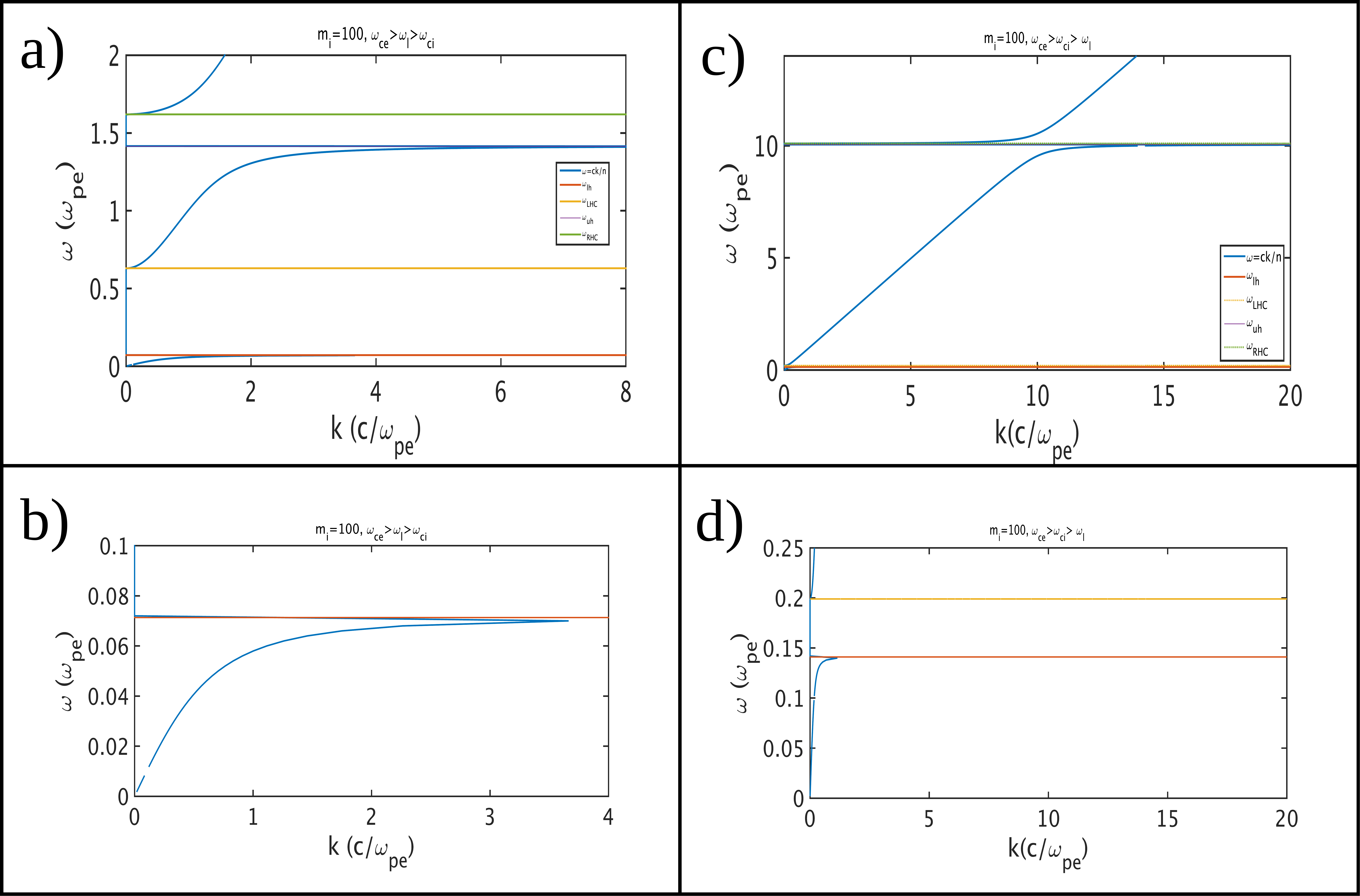}

		\caption{Dispersion curve for two criteria(criteria I $(\omega_{ce}>\omega_l>\omega_{ci})$ and criteria II $\omega_{ce}>\omega_{ci}>\omega_{l})$) are shown in subplot (a) and (c)  respectively). In subplot (b) and (d), we expand the dispersion curves shown in (a) and (c) respectively. From these curves, we can observe that while criteria I has two stop bands and three pass bands, criteria II dispersion has $\omega=ck$ all along. }
		
		\label{xmode_compare}
		
	\end{figure*}

	\begin{figure*}
		\includegraphics[width=\textwidth]{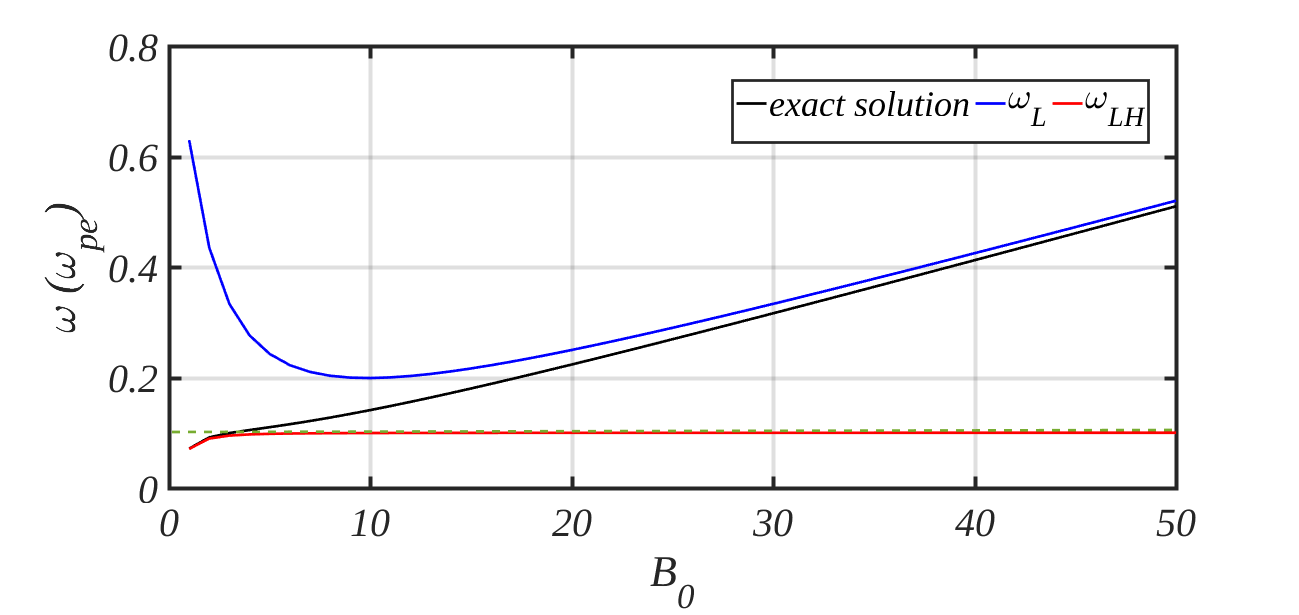}

		\caption{Comparison of exact solution for $\omega$ with cut off frequency and analytical expression of LH as function of $B_0$. As one can observe that at lower magnetic field value there's gap between values of $\omega_{L}$ and $\omega_{LH}$ but at high value of magnetic field where exact solution of $\omega$ will play a role it matches with $\omega_{L}$ signifying the vanishing of stop band. Gray dash-line indicates the $\omega_{pi}$ position in the plot.}
		
		\label{analytic1}
		
	\end{figure*}
	
\begin{figure*}
		\includegraphics[width=\textwidth]{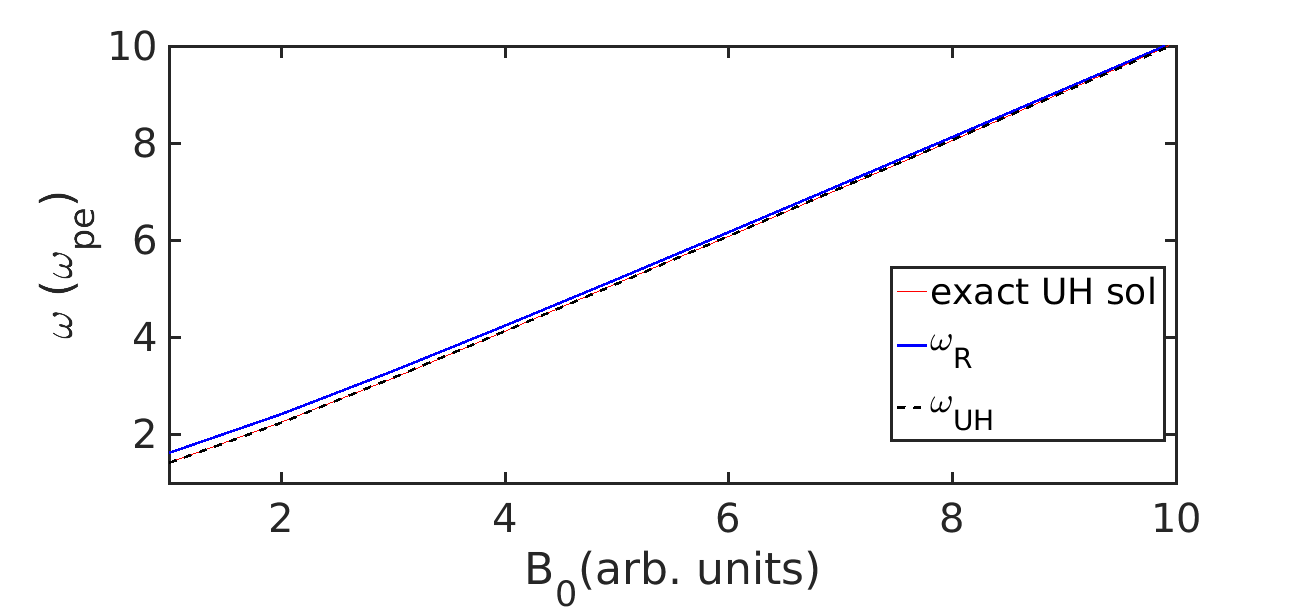}

		\caption{Comparison of one exact solution for $\omega$ with cut off frequency and analytical expression of UH as function of $B_0$. As one can observe that at lower magnetic field value there's gap between values of $\omega_{R}$ and $\omega_{UH}$ but at high value of magnetic field where exact solution of $\omega$ will play a role it matches with $\omega_{R}$ indicative to the vanishing of stop band.}
		
		\label{analytic2}
		
	\end{figure*}
	\begin{figure*}
		\includegraphics[width=\textwidth]{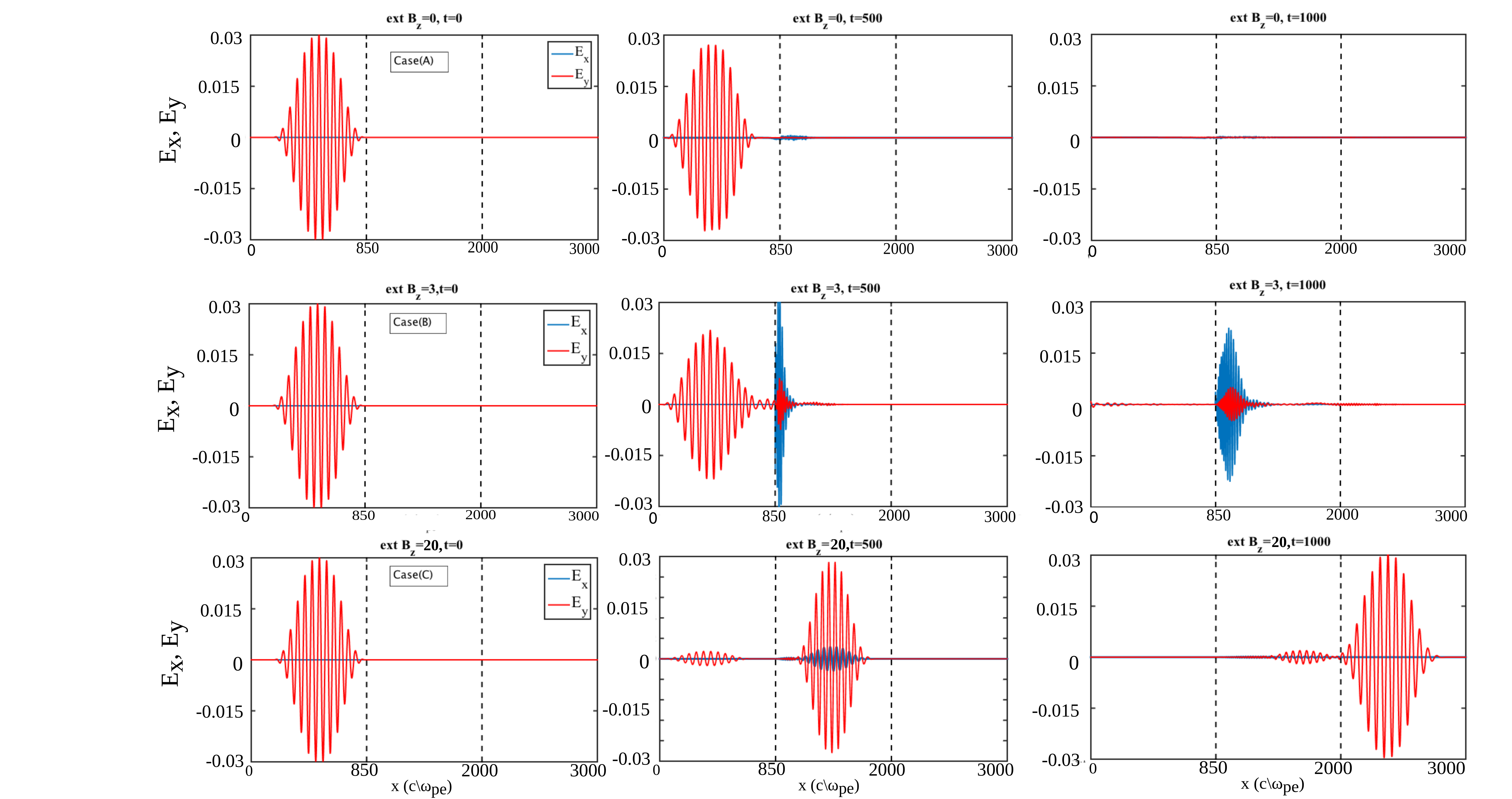}

		\caption{Spatial variation of $E_x$ and $E_y$ for the three cases at different times. Initially, only $E_y$(due to laser field) is present in the system. In case (A), laser gets reflected from the plasma boundary without being able to interact with it. On the other hand, laser is able to interact with plasma in Case(B) and we observe generation of $E_x$ in the system. In case (C), medium becomes transparent to laser and laser just passes through the plasma medium unhindered.}
		
		\label{ExEy}
		
	\end{figure*}

	\begin{figure*}
		\includegraphics[width=\textwidth]{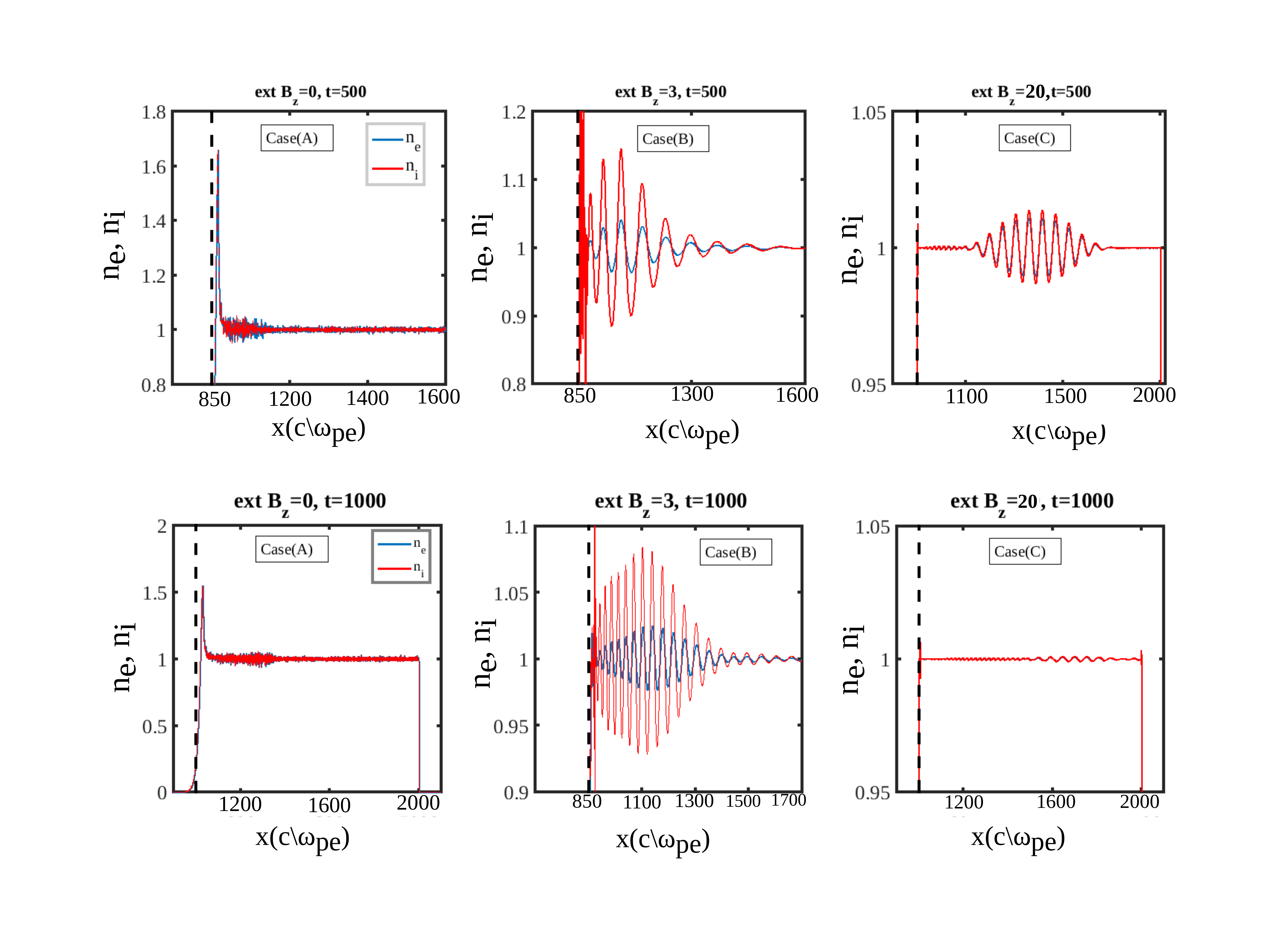}

		\caption{Spatial variation of number density of electrons and ions for the three cases at different times showing that plasma density gets perturbed at the interface for case(A) whereas for case(B), we observe density fluctuations in the bulk plasma as well which retain themselves even after laser has moved out of the simulation box($t=1000$).For case(C), on the other hand, we observe some density perturbations in the bulk plasma at the time of interaction with laser($t=500$) but these fluctuations are not retained by plasma after the interaction with laser is over.}
		
		\label{neni}
		
	\end{figure*}

\begin{figure*}
	\includegraphics[width=\linewidth]{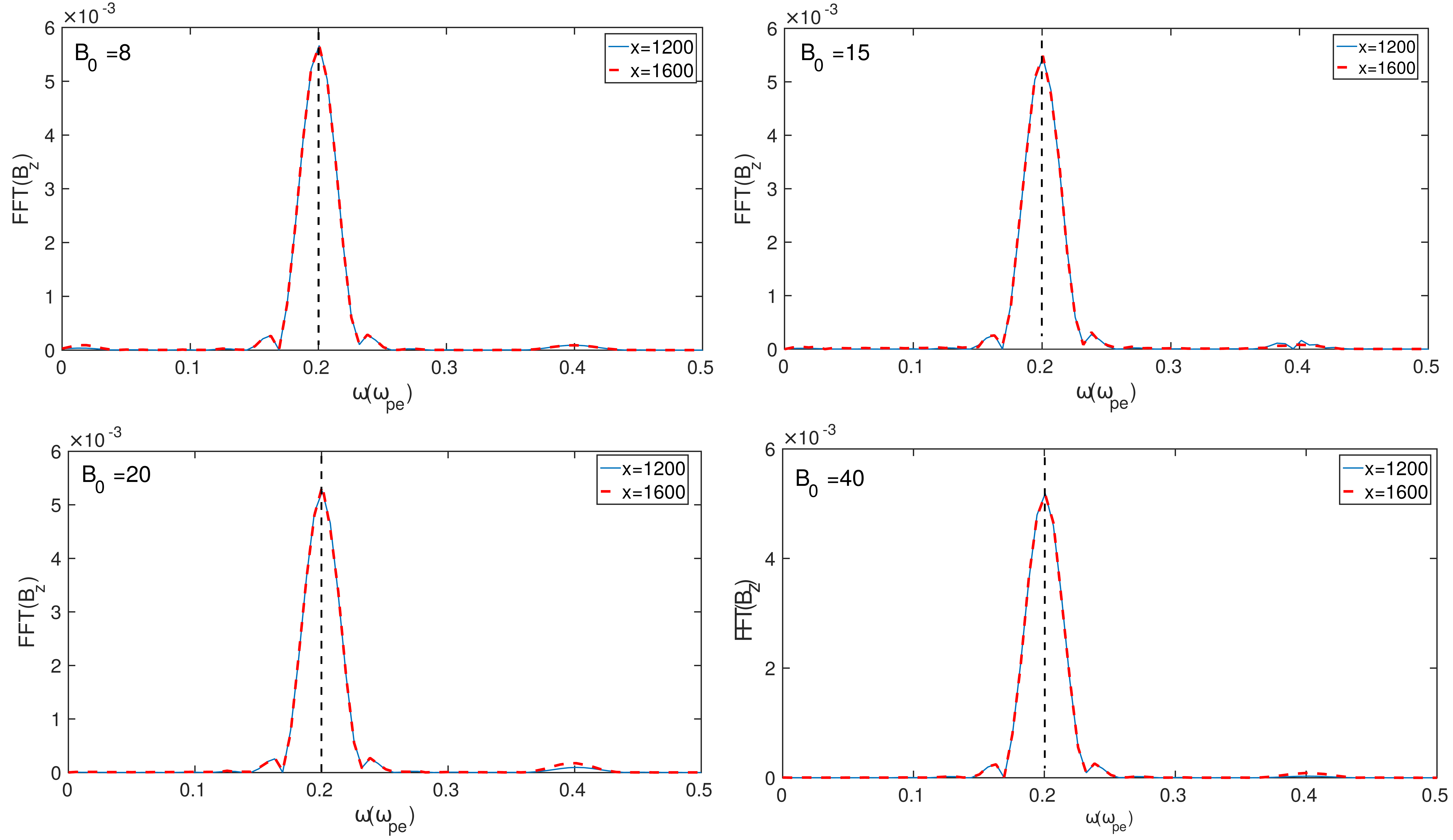}

	\caption{FFT of $B_{z}$ with time at two different locations in bulk plasma, showing that the peak frequency of electromagnetic wave does not change and it propagates unhindered in the plasma medium for different magnetic fields($B_0=8, 15, 20, 40$).}
	
	\label{fft_t}
	
\end{figure*}
   \begin{figure*}
   	\includegraphics[width=\linewidth]{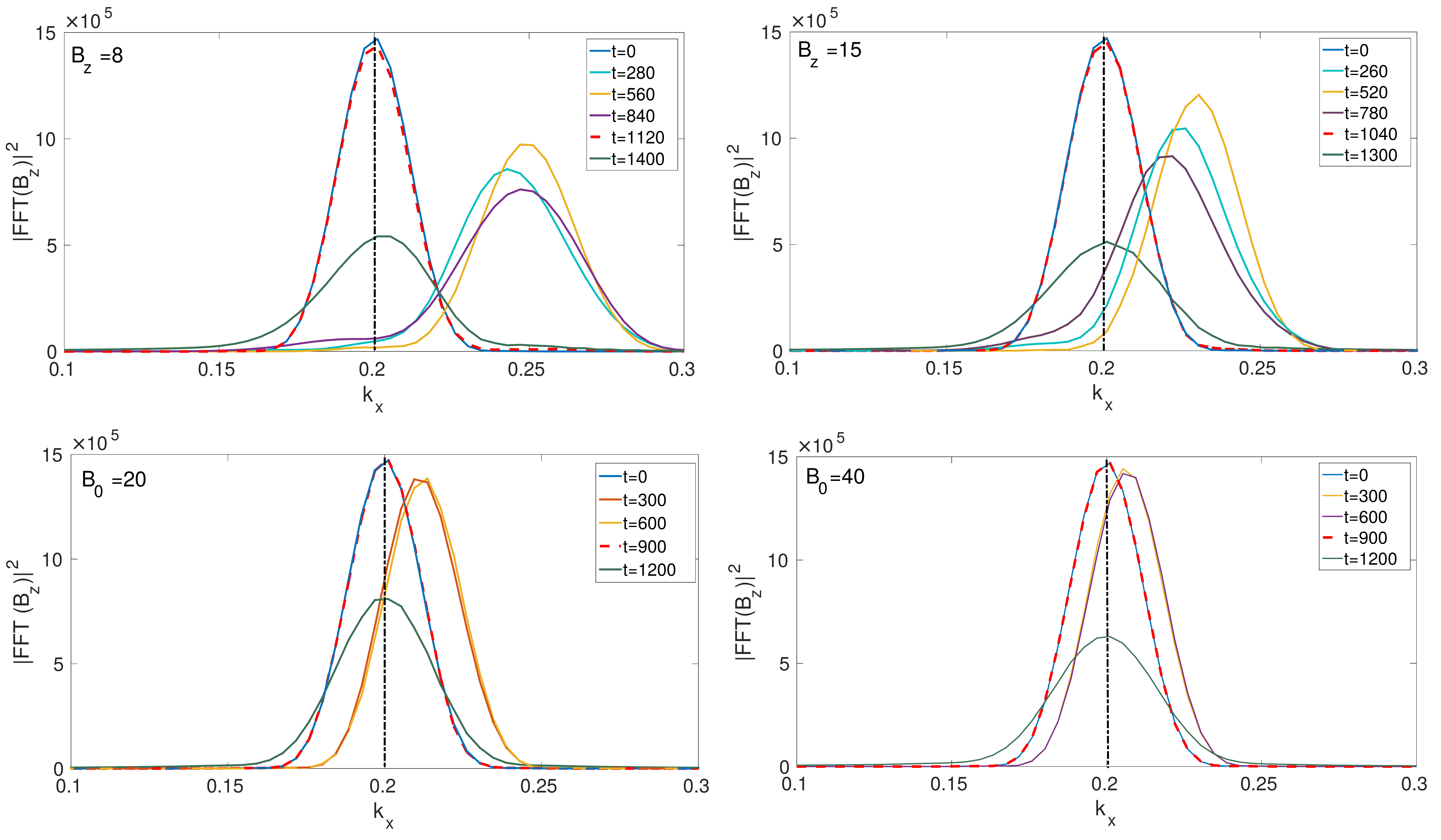}

   	\caption{FFT of $B_{z}$ with x at different times showing that as laser interacts with plasma, $k_x$ shifts to a larger value than the initial $k_x$. This shift reduces as we increase the magnitude of applied magnetic field(shown in different subplots). This shift in $k_x$ is also reflected in velocity of EM wave approaching c in the medium (Table \ref{v_table}) and the transmission coefficient approaching unity (Table \ref{coefficient_table}). }
   	
   	\label{fft_x}
   	
   \end{figure*}
   
   \begin{figure*}
		\includegraphics[width=\textwidth]{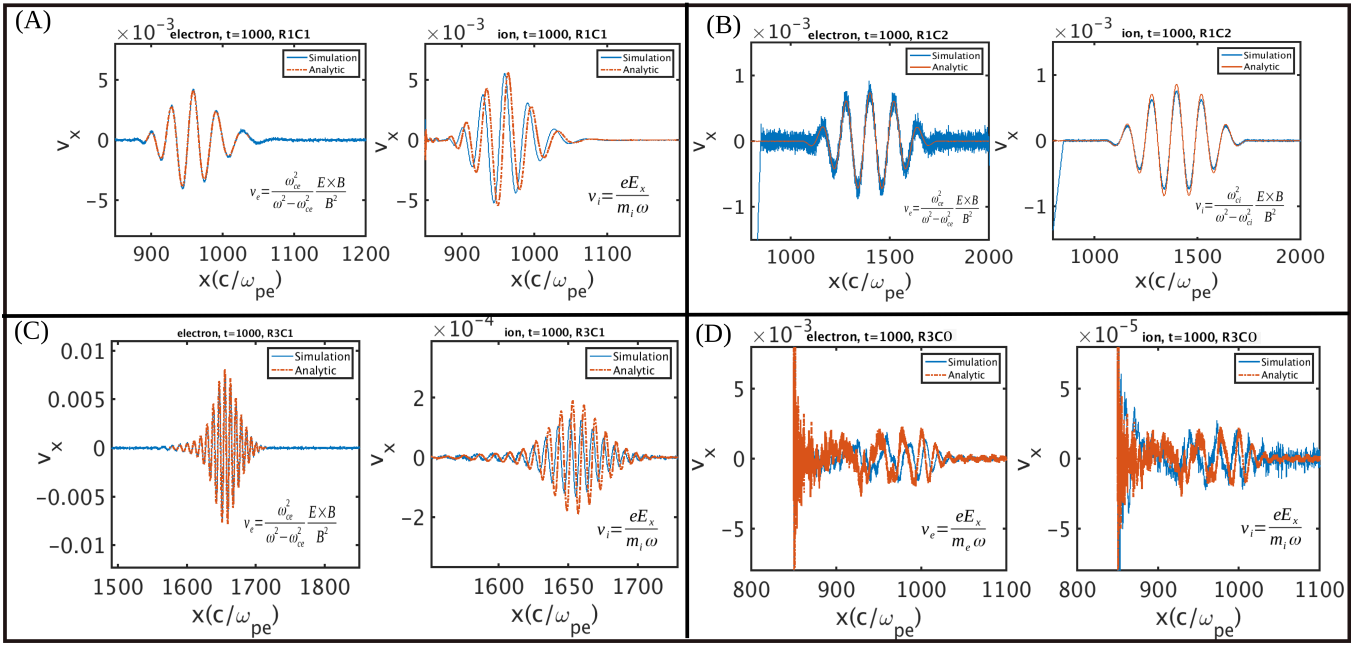}

		\caption{Comparison of electron and ion drift theoretically(eq. \ref{EXB}) and simulation results.  In criteria II both species are magnetised and their drift velocity in longitudinal direction is also same such that there's no charge separation which can inhibit the propagation of EM wave (B). In the case when $\omega_l>\omega_{ce}$, there's generation of electrostatic perturbation which helps in coupling laser energy into plasma irreversibly shown in subplot(D).  For criteria I in region I and III, ions are un-magnetised while electrons are tightly bound to magnetic field(in subplot(A) and (C)).   }
		
		\label{comparison}
		
	\end{figure*}
   
  \end{widetext} 
  \end{document}